\documentclass[12pt,a4paper]{article}      
\usepackage{isolatin1,a4,epsf,amssymb}          

\author{J. Diemand, Ch. Mathys and D. Wyler \\
Institute of Theoretical Physics\\
University of Zürich}

\title{Dynamical Instabilities of Brane World Models}

\begin{document}

\maketitle

\begin{abstract}
We show the instability of two self-tuning brane world models with gauge invariant linear 
perturbation theory. This general method confirms a known instability of the original model with 
vanishing bulk potential. We also show the dynamical instability of a recently proposed 
self-tuning "smooth" brane model and its limit, the Randall-Sundrum
model. Astonishingly, we also find instability under purely gravitational 
perturbations.
\end{abstract}

\section{Introduction}
Superstring theories and 
especially M--theory have revived the idea that the universe is a 
$3+1$ dimensional hyper-surface, a $3$--brane, of a $10$ or $11$
dimensional spacetime. This fundamental spacetime could be 
a product of a four dimensional Lorentz manifold with an
$n$ dimensional compact space of volume $V_n$. Then, the effective four 
dimensional Planck mass, 
$M_{\mbox{\tiny{eff}}}\equiv \sqrt{1/(8\pi G_N)} \simeq  2.4 \times
10^{18}$ GeV is related by 
\begin{equation}
M_{\mbox{\tiny{eff}}}^2 = M_{\mbox{\tiny{f}}}^{n+2} V_n.
\end{equation}
to the $4+n$ dimensional
fundamental Planck mass, $M_{\mbox{\tiny{f}}}$. Since 
$M_{\mbox{\tiny{f}}}$ is much smaller than
$M_{\mbox{\tiny{eff}}}$ and may even be close to the electro-weak
scale the long-standing hierarchy problem~\cite{arkani-hamed} might be resolved. 
Clearly, this is an intriguing idea and it has attracted a lot of attention.

Instead of a factorizable space-time with a compact extra-dimensional 
space, Randall and Sundrum~\cite{randall}(RS, see also (\cite{frolich})), 
proposed a five 
dimensional model, in 
which the metric on the $3$-brane is multiplied by 
an exponentially decreasing `warp' factor such that transverse lengths
become small already at short distances along the fifth
dimension. The brane is embedded in an Anti-de Sitter space, 
but unfortunately, a fine-tuning relation
between the brane tension and the negative cosmological constant in 
the bulk has to be satisfied. 

Another possibility was discovered by  Arkani-Hamed, Dimopuolos, Kaloper and 
Sundrum \cite{adks} and by Kachru, Schulz and Silverstein \cite{kss}
(ADKS-KSS). In this model, there is a 4D-Poincar\'e invariant solution 
for each choice of the input parameters which is stable with 
respect to radiative corrections of the brane tension. This feature is 
called ``self-tuning'' and would explain why we observe a vanishing 
cosmological constant, i.e. it would solve the cosmological constant problem \cite{straumann}. 
However, the model has singularities in the bulk
which seem to be inevitable in self-tuning models of this kind \cite{csaki}. 
It has also turned out that the
model needs a fine tuning for the regularization of the
singularity in the bulk \cite{nilles}. A variant of these ideas was recently 
given by 
Kehagias and Tamvakis who introduce a self-tuning 'smooth' brane model 
without singularities \cite{tamvakis1}, \cite{tamvakis} (KT). 

These models are static. This immediately raises the question
whether their essential features persist in time. For the ADKS-KSS model this is 
not the case: 
Arbitrarily close to the static self-tuning solution there are time-dependent 
solutions, 
which lead to an expansion or a contraction of the brane world; the model is 
therefore dynamically unstable \cite{binetruy}. 

Using gauge invariant linear perturbation theory 
we show in this paper that the ADKS-KSS model and the KT model are 
dynamically unstable against small 
variations in the scalar field and even homogeneous, 
purely gravitational perturbations. The instability of the KT model also 
implies the instability of the RS model because it is a 
certain 'brane limit' of 
the KT model. This limit does not modify 
the purely gravitational mode; therefore the RS model is also 
unstable under this mode. This result was also found in \cite{durrer}. 

In section \ref{section2} we write down the dynamical equations for the 
gravitational and the scalar field which couples to gravity and to the brane 
tension. We derive linear 
perturbation equations which describe the evolution of small deviations 
from a static solution. In section \ref{pert} we give the perturbation 
equations in gauge invariant form 
and prove the instability of the KT model and the ADKS-KSS model. 
In the last section we present our results and conclusions.

\section{The Linearized Einstein Equations}\label{section2}

We consider a general five-dimensional action\footnote{{\bf Conventions and Notation:} We use a mostly positive Lorentzian signature $(-+\ldots +)$ and the definition of the curvature in terms of the metric is such that a Euclidean sphere has  positive curvature. We write the 5D Einstein equations as $G_{MN}=\kappa^2 T_{MN}$ with the constant $\kappa$ related to the 5D Planck mass $\kappa^2=1/4M_5^3$. Bulk indices are denoted by capital Latin indices and brane indices by Greek indices: $x^\mu$ are coordinates on the brane ($t$ is the time coordinate and $x^i$ the spatial ones), and $y$ is the coordinate along the fifth dimension such that the brane is located at $y=0$.} with a bulk scalar field $\phi$ with potential $U(\phi)$ and one brane with brane tension 
$\lambda(\phi)$, 
\begin{equation}  \label{action}
 {\cal S}=\int
d^5x\sqrt{|g_5|}\left(\frac{1}{2\kappa^2} R-\frac{1}{2}(\partial \phi)^2
-U(\phi)\right) -
\int d^{\,4}x\,\sqrt{|g_4|}\lambda(\phi_b)\, ,
\end{equation}
where $g_5$ and $g_4$ are, respectively, the determinants of the 5D metric $g_{MN}$ and the 4D metric induced on the brane, $g_{\mu\nu}$, and 
$\phi_b$ denotes the value of the scalar field on the brane.
The tension of the brane couples to the bulk scalar field in a way defined by the function $\lambda(\phi)$. The Einstein and scalar field equations are
\begin{eqnarray} \label{einsteineq}
&&
G_{MN} = \kappa^2\left( \partial_M\phi \partial_N\phi
-g_{MN}\left( \frac{1}{2} (\partial \phi)^2 + U(\phi) \right)\right)
-\frac{\lambda(\phi)}{2} g_{\mu\nu} \delta^\mu_{M} \delta^\nu_{N}
\frac{\sqrt{|g_4|}}{\sqrt{|g_5|}}\, \delta (y) \; , \nonumber \\
&& \label{phiequation}
 \frac{1}{\sqrt{|g_5|}} \partial_M \left(\sqrt{|g_5|}g^{MN}\partial_N
\phi\right)
-\frac{1}{2 \kappa^2} \frac{d \lambda}{d\phi} 
\frac{\sqrt{|g_4|}}{\sqrt{|g_5|}}\, \delta (y) = \frac{\partial U(\phi)}{\partial \phi} \ .
\end{eqnarray}
To investigate the dynamical behaviour of brane world models, we use the following ansatz for the metric:
\begin{equation} \label{dynmetric}
        g= e^{2A(t,y)} \eta _{\mu\nu} dx^{\mu} dx^{\nu} + b(t,y)^2dy^2.
\end{equation} 
Inserting this into the metric, one finds in the bulk ($y\not= 0$) 
the Einstein equations
( $'$ denotes $\partial_y$ and $\dot{}$ denotes $\partial_t$)
\begin{eqnarray}\label{einsteineq-00}
        3e^{-2A}(\dot{A}^2 b^2  + \dot{A}\dot{b} b)  - 6 A'^2
        &-& 3A'' + 3\frac{A'b'}{b} \nonumber \\ 
        &=& \kappa^2 \left(\frac{1}{2} \dot{\phi}^2 b^2e^{-2A} + \frac{1}{2} \phi'^2  + b^2 U(\phi)  \right)  \\ 
        (-2\ddot{A} b^2 - \dot{A}^2 b^2 -\dot{A}\dot{b} b &-& \ddot{b}b ) e^{-2A} \label{einsteineq-ii}
        +  6 A'^2 + 3A'' - 3 A'\frac{b'}{b} \nonumber \\       
        &=& \kappa^2 \left(\frac{1}{2} \dot{\phi}^2 b^2e^{-2A} - \frac{1}{2} \phi'^2  -b^2 U(\phi) \right)  \\   
        -3 b^2 e^{-2A}( \ddot{A} + \dot{A}^2 ) + 6 A'^2  \label{einsteineq-44}
        &=& \kappa^2 \left(\frac{1}{2} \phi'^2 + \frac{b^2 e^{-2A}}{2} \dot{\phi}^2 -b^2 U(\phi)  \right) \\
        -3\dot{A}' + 3 A'\frac{\dot{b}}{b}   
        &=& \kappa^2 \dot{\phi}\phi'  \; , \label{einsteineq-04}
\end{eqnarray}
and the scalar field equation 
\begin{equation} \label{phieq}
e^{2A} b \left( -\dot{b}\dot{\phi} - b\ddot{\phi} - 2 \dot{A}b \dot{\phi} \right)
 + 4 A' \phi'  - \frac{b' \phi'}{b} + \phi'' = 
b^2 \frac{\partial U}{\partial \phi} \; .
\end{equation}
The $\delta(y)$-terms in the action lead to jump conditions for the derivatives of the fields at the brane
\begin{eqnarray}\label{jumpcond-A}
        \left[ A(t,y)'  \right]_{y=0^-}^{y=0^+}& = &- \frac{\lambda(\phi_b(t))}{6} \\
        \left[ \phi(t,y)'  \right]_{y=0^-}^{y=0^+}
        & = &\frac{1}{2\kappa^2} \frac{d\lambda(\phi_b(t))}{d \phi}\; . \label{jumpcond-phi}
\end{eqnarray}
To linearize these equations around a static solution ($\phi_0(y)$, $A_0(y)$, and $b=1$)\footnote{We demand $b(y,t)=$ constant, so with a redefinition of the $y$-coordinate we get $b=1$.}, we set 
\begin{eqnarray} 
        \phi(t,y)&=&\phi_0(y)+\delta\phi(t,y) \nonumber\\
        A(t,y) &=& A_0(y) + \delta A(t,y)  \qquad \mbox{and} \qquad
        b(t,y) = 1 + \delta b(t,y) \;.
\end{eqnarray}
The static solution has to satisfy the zeroth 
order field equations in the bulk, which reduce to
\begin{eqnarray}
6 A'^2  &=& \kappa^2 \frac{1}{2}(\phi')^2-U(\phi) \; ,\label{field-e1-0} \\
-3 A''- 6A'^2 &=& \kappa^2\frac{1}{2}(\phi')^2+U(\phi) \; ,\label{field-e2-0}\\
4A'\phi'+\phi''&=&\frac{\partial U}{\partial \phi}\;,\label{phieq-0}
\end{eqnarray}
and the zeroth order jump conditions at the brane
\begin{eqnarray}\label{jumpcond-A-0}
        \left[ A_0'  \right]_{y=0^-}^{y=0^+}& = &- \frac{\lambda(\phi_{0b})}{6} \\
        \left[ \phi_0'  \right]_{y=0^-}^{y=0^+}
        & = &\frac{1}{2\kappa^2} \frac{d\lambda(\phi_{0b})}{d \phi}\; . \label{jumpcond-phi-0}
\end{eqnarray}
Many solutions of these equations are known; see \cite{gubser} for a general
solving procedure for arbitrary $U(\phi)$. 

%We will later discuss the stability of two self-tuning solutions, the 
%Kehagias-Tamvakis (KT) model \cite{tamvakis1},\cite{tamvakis} and the 
%ADKS-KSS model \cite{adks},\cite{kss}.
Now if we keep terms up to first order in $\delta$, the (00), (ii), (44) and (04) components of the Einstein equations are
\begin{eqnarray}\label{einsteineq-00-1}
         - 12 A_0'\,\delta A -  3\,\delta A'' &+& 3A_0'\,\delta b' \nonumber \\     
        &=& \kappa^2 \left( \phi_0'\delta\phi' +   \frac{\partial U(\phi_0)}{\partial\phi}\delta\phi 
+2 U(\phi_0) \delta b\right)   \\ 
        (-2  \,\ddot{\delta A} &-&  \, \ddot{\delta b}) e^{-2A_0}
        + 12 A_0'\,\delta A + 3\,\delta A'' - 3A_0'\,\delta b' \nonumber \\       
        &=& \kappa^2 \left( - \phi_0'\delta\phi' -   \frac{\partial U(\phi_0)}{\partial\phi}\delta\phi 
-2 U(\phi_0) \delta b\right) \label{einsteineq-ii-1}   \\   
        -3   e^{-2A_0}\,\ddot{\delta A} &+& 12 A_0'\,\delta A'   \nonumber \\ 
        &=& \kappa^2 \left( \phi_0'\delta\phi' -   \frac{\partial U(\phi_0)}{\partial\phi}\delta\phi 
- 2 U(\phi_0) \delta b\right) \label{einsteineq-44-1}  \\
        -3\,\dot{\delta A}' + 3 A_0'\,\dot{\delta b}   
        &=& \kappa^2 \phi'\,\dot{\delta\phi}  \; , \label{einsteineq-04-1}
\end{eqnarray}
and the linearized scalar field equation is
\begin{eqnarray} \label{phieq-1}
        \delta \phi '' &=&   e^{-2A_0}\,\ddot{\delta\phi } -4 \phi_0'\delta A'
+ 2   \frac{\partial U(\phi_0)}{\partial \phi}\,  \delta b \nonumber \\  
&&+  \frac{\partial^2 U(\phi_0)}{\partial^2 \phi}\, \delta \phi  
  - 4A_0'\, \delta\phi ' + \phi_0' \,\delta b ' \; .
\end{eqnarray}
At the brane the perturbation amplitudes have to fulfill
 the first order jump conditions
\begin{eqnarray}\label{jumpcond-A-1}
         \left[ \delta A'  \right]_{y=0^-}^{y=0^+}& = &- \frac{1}{6}\frac{d\lambda(\phi_0)}{d\phi} \,\delta\phi \\
        \left[ \delta\phi'  \right]_{y=0^-}^{y=0^+}
        & = &\frac{1}{2\kappa^2} \frac{d^2\lambda(\phi_0)}{d \phi^2} \,\delta\phi \;.     \label{jumpcond-phi-1}  \; 
\end{eqnarray}

\section{Perturbation Calculations}\label{pert}

The ideas and methods in this section come from cosmological perturbation 
theory (for an introduction, see \cite{cpt1}), originally developed to describe
the formation and evolution of density fluctuations; here we use them
to study questions of stability.
Similar methods were used in \cite{cline} to derive the Friedmann 
equations of a brane world model.

General relativity is invariant under diffeomorphisms (coordinate 
transformations). 
This means that for an arbitrary diffeomorphism $\varphi$ and its 
pullback $\varphi^*$, 
the metrics $g$ and $\varphi^* (g)$ describe the same geometry.
We look at small perturbations $\delta g$ 
around  a static background metric $g_0$ and allow only 
diffeomorphisms which leave the background metric invariant. 
These 'infinitesimal' transformations can be described as the flow of 
an 'infinitesimal' vector field $X$ and the perturbation of an arbitrary 
tensor field $Q=Q_0 +\delta Q$ 
obeys the gauge transformation law
\begin{equation} \label{trafo}
\delta Q \rightarrow \delta Q + L_X Q_0   \;, 
\end{equation}
where $L_X$ denotes the Lie derivative in the direction of $X$. 
Taking the vector field to be of 'scalar' type, i.e. 
\begin{equation}
        X=X^{\mu} \partial_{\mu}+X^4 \partial_4\;, \qquad X^{\mu} = \eta ^{\mu\nu} \xi_{,\nu}\;,
\end{equation}
with an arbitrary scalar field $\xi(x^M)$ and noting that the most general 
scalar\footnote{We only need perturbations of 
the scalar type because in $T_{\mu \nu}$ we have just a scalar perturbation 
of the scalar field $\phi$.} perturbations of the metric can be written as ( $_{,M}$ 
denotes $\frac{\partial}{\partial x^M}$ )
\begin{equation}\label{pertansatz}
\delta g= 2e^{2A_0} \left( \delta A \,\eta_{\mu\nu} 
+\delta E_{,\mu\nu}\right) dx^{\mu} dx^{\nu} 
+ 2 \,\delta b\, dy^2 + 2 \,\ \delta B_{,\mu}\,dx^{\mu} dy,
\end{equation}
we arrive at
\begin{eqnarray}
        \delta A &\rightarrow & \delta A + A_0' X^{4} \label{trafodA} \;,\\
        \delta b &\rightarrow & \delta b +   X^{4}_{,4}  \;,\\
\delta B_{,\mu} &\rightarrow & \delta B_{,\mu} +   X^4_{,\mu}+ e^{2A_0}\xi_{,\mu 4} \nonumber\\
\Rightarrow \delta B  &\rightarrow & \delta B  +    X^{4}+ e^{2A_0}\xi_{,4}  \label{trafodB} \;,\\
\delta E_{,\mu\nu} &\rightarrow & \delta E_{,\mu\nu} + \xi_{,\mu \nu} \nonumber\\
\Rightarrow \delta E  &\rightarrow & \delta E  +\xi \label{trafodE} \;.
\end{eqnarray}
In brane world models there is a jump in $A_0'$ at the brane, therefore (\ref{trafodA}) implies $X^4(x^{\mu},y=0)=0$ and $\delta A$ restricted to the brane, $\delta A_b$, is gauge invariant.

The scalar field perturbation $\delta\phi$ is defined by $\phi(t,y)=\phi_0(y)+\delta\phi(t,y)$; according to (\ref{trafo}) one has
\begin{equation}
 \delta \phi \rightarrow  \delta \phi + \phi_0' X^{4} \; ,
\end{equation}
and again the restriction to the brane, $\delta \phi_b$, is gauge invariant.

The gauge invariant terms for the perturbation amplitudes then are \footnote{The first two terms, $\delta A_L$ and $\delta b_L$, correspond to the Bardeen potentials in cosmological perturbation theory if we identify our extra dimension with the time dimension and our 4D Minkowski space with the 3D space of constant curvature.}
\begin{eqnarray}
        \delta A_L &:=& \delta A + A_0'\left(e^{2A_0}\,\delta E_{,4}- \delta B \right)\\
        \delta b_L &:=& \delta b + \left(e^{2A_0}\,\delta E_{,4}- \delta B \right)_{,4}  \\
\delta \phi_L &:=& \delta \phi + \phi_0' \left(e^{2A_0}\,\delta E_{,4}- \delta B \right)  \; .
\end{eqnarray}
Because of (\ref{trafodB}) and (\ref{trafodE}), it is always possible to 
find a gauge with $\delta B =\delta E = 0$ (longitudinal gauge). In this gauge, the metric is diagonal and has the form (\ref{dynmetric}), therefore it is much simpler to calculate the first order Einstein equations in the longitudinal gauge (\ref{einsteineq-00-1})-(\ref{einsteineq-04-1}) and later rewrite them in a gauge invariant form. In the longitudinal gauge each of the gauge invariant amplitudes above reduces to the corresponding gauge dependent term, for example $\delta A_L =\delta A $ etc. 

Two other gauge invariant combinations of the perturbation amplitudes will be useful in the following:
\begin{eqnarray}
        \Psi &:=& -\delta A' + A_0' \,\delta b - \frac{\kappa^2}{3} \phi_0' \,\delta\phi \label{Psi} \\
        \Phi &:=& \phi_0 '' \,\delta \phi - \phi_0' \,\delta \phi' + \phi_0'^2 \,\delta b \; ,\label{Phi}
\end{eqnarray}
where the gauge invariance of $\Psi$ is a consequence of the zeroth order 
equation $3 A_0''= -\kappa^2 \phi_0'^2 $, which follows from (\ref{field-e1-0}) and (\ref{field-e2-0}). In the definitions (\ref{Psi}) and (\ref{Phi}) we could of course equivalently use the gauge invariant amplitudes $\delta A_L$, $\delta b_L$ and $\delta \phi_L$; the $\delta B$ and $\delta E$ terms cancel.

Using the zeroth order equations (\ref{field-e1-0})-(\ref{phieq-0}) we can absorb the terms with the potential $U(\phi_0)$ in the first order equations into combinations of the gauge invariant terms $\Psi$ and $\Phi$. In gauge invariant form, the (00), (00) + (ii), (44) and (04) components of the first order Einstein equations (\ref{einsteineq-00-1})-(\ref{einsteineq-04-1}) are 
\begin{eqnarray}\label{gi-einsteq-00}
        4 A_0'\Psi + \Psi' & = & 0 \\
\left(-2 \,\ddot{\delta A_L} - \ddot{\delta b_L} \right) e^{-2A_0}   
&=& 0 \label{gi-einsteq-00ii} \\
        -3e^{-2A_0}  \,\ddot{\delta A_L} -12 A_0' \Psi + \kappa^2 \Phi &=& 0 \label{gi-einsteq-44} \\
        \dot{\Psi} & = & 0 \; ,\label{gi-einsteq-04}
\end{eqnarray}
and (\ref{phieq-1}), written in gauge invariant form, gives 
\begin{eqnarray} \label{gi-phieq}
        \delta \phi_L'' &=&   e^{-2A_0}\,\ddot{\delta\phi_L} + 4 \phi_0'\Psi
+ \left(2   \frac{\partial U_0(\phi_0)}{\partial \phi} - 4 \phi_0' A_0' \right) \delta b_L\nonumber \\  
&&+\left(  \frac{\partial^2 U_0(\phi_0)}{\partial^2 \phi} + \frac{4\kappa^2}{3} \phi_0'^2  \right) \delta \phi_L 
  - 4A_0'\, \delta\phi_L' + \phi_0' \,\delta b_L' \; .
\end{eqnarray} 
The first order jump conditions (\ref{jumpcond-A-1}) and (\ref{jumpcond-phi-1}) are already gauge invariant because of $X^4(x^{\mu},y=0)=0$ and $X^4 \in C^1$.

The procedure to find suitable perturbation amplitudes $\delta A_L$, $\delta b_L$ and 
$\delta \phi_L$ is to solve (\ref{gi-einsteq-00})-(\ref{gi-phieq}) in the bulk, i.e. for $y<0$ and $y>0$, 
and then glue the solutions continuously together at $y=0$ and check 
whether they satisfy the 
jump conditions. 

\subsection{A Solution with $\delta b_L=0$}

Now we derive a solution to the bulk equations 
which we will later use to 
prove the instability of the KT and the ADKS-KSS self-tuning model.
Equations (\ref{gi-einsteq-00}) and (\ref{gi-einsteq-04}) imply
\begin{equation}
\Psi(t,y)= K_1 e^{-4A_0(y)} \; ,
\end{equation}
where $K_1$ is an integration constant. We try to find perturbations linear in $t$, and to simplify matters we set $\delta b_L \equiv 0$. (\ref{gi-einsteq-44}) then yields
\begin{equation}
\Phi = \frac{12}{\kappa^2}A_0'\Psi =\frac{12}{\kappa^2}A_0'  K_1 e^{-4A_0(y)} \; .
\end{equation}
According to the definitions of $\Phi$ and $\Psi$ (\ref{Psi})-(\ref{Phi}), 
this implies
\begin{equation}\label{deltaphik'}
\delta \phi'_L = \frac{\phi_0''}{\phi_0'} \delta\phi_L - \frac{12}{\kappa^2} \frac{A_0'}{\phi_0'} K_1 e^{-4A_0} \; .
\end{equation}
The subsequent integration gives
\begin{equation}\label{deltaphik}
\delta \phi_L = -K_1 \phi_0' \int \frac{4A_0'}{\phi_0'^2} e^{-4A_0} \, dy + k_2(t) \phi_0' \; ,
\end{equation}
with another (time dependent) integration constant $k_2$. From the definition of $\Psi$ we can find $\delta A'_L$
and then the perturbation amplitude $\delta A_L$ must be
\begin{equation}\label{deltaAk}
\delta A_L = K_1 \int \phi_0'^2 \left( \int \frac{4A_0'}{\phi_0'^2} e^{-4A_0} \, dy \right) dy 
+ A_0' k_2(t) - K_1\int e^{-4A_0}\, dy + k_3(t)  \; .
\end{equation}
One can check that the scalar field equation is also satisfied.

Since the solutions were assumed to be linear in $t$, the integration 
constants $k_2(t)$ and $k_3(t)$ have the form
$k_i(t)=K_i + h_i t $, and we get a five-parameter family of perturbation amplitudes
\begin{eqnarray}
\delta A_L & =& K_1 \int \phi_0'^2 \left( \int \frac{4A_0'}{\phi_0'^2} e^{-4A_0} \, dy \right) dy \nonumber \\
&&+ A_0' K_2 + A_0' h_2 t - K_1\int e^{-4A_0}\, dy + K_3 + h_3 t  \label{deltaA} \\
        \delta \phi_L &=& -K_1 \phi_0' \int \frac{4A_0'}{\phi_0'^2} e^{-4A_0} \, dy 
+ \phi_0' K_2 + \phi_0' h_2 t \label{deltaphi} \; .
\end{eqnarray}
We are only interested in time-dependent perturbations, so we set $K_i=0$
and obtain
\begin{eqnarray}
\delta A_L & =&  A_0' h_2 t  + h_3 t  \label{deltaA1p} \\
        \delta \phi_L &=&  \phi_0' h_2 t \label{deltaphi1p} \; .
\end{eqnarray}

Note that for $h_i=0$ all the gauge invariant terms are zero, and the solution is equivalent to the unperturbed, static solution $A=A_0$, $b=1$, $\phi=\phi_0$ and $\delta B= \delta E =0$.

\subsection{Instabilities in the KT-Self-Tuning-Model}

In the self-tuning model proposed by Kehagias and Tamvakis there is no brane term proportional to $\delta(y)$ in the five dimensional action (\ref{action}), but a 'smooth' brane is dynamically formed by the scalar field coupled to the 
Standard Model. So we have to set $\lambda(\phi)\equiv 0$ to study the stability of this model. Now the field equations are valid everywhere and there are no jumps in the derivatives $A(t,y)$ and $\phi(t,y)$, the jump conditions are trivial. The potential is chosen to be
\begin{equation}
U(\phi)=V_0\left(1-k^2\sin^2(\beta\phi)\right)  \qquad \mbox{with} 
\qquad k^2=1+\frac{4\kappa^2}{3\beta^2} \;.
\end{equation}
The static solution $\phi_0(y)$ has the kink-like form
\begin{equation}
\phi_0 =\frac{2}{\beta}\arctan\Big{(}\tanh(ay/2)\Big{)} \, , 
\label{tamvakis-phi}
\end{equation}
where the parameters are related and restricted by
\begin{equation}
a^2\equiv 2V_0\beta^2=\frac{2V_0}{3M^3(k^2-1)} 
\qquad \mbox{and} \qquad k^2 > 1 \; ,
\end{equation}
and the static warp function $A_0(y)$ is
\begin{equation}
A_0 (y)=\frac{(1-k^2)}{4}\ln\cosh(ay)\, .\label{tamvakis-A} 
\end{equation}
For this solution, the four-dimensional Planck mass has a finite value
\begin{equation}\label{tamvakis-planck} 
M_p^2 = M_5^3 \int_{-\infty}^{\infty} e^{2A_0(y)}\,  d y  =
\frac{4 M_5^3 \sqrt{\pi}}{a(k^2-1)}\frac{\Gamma\left((k^2+3)/4)\right)}{\Gamma\left((k^2+1)/4\right)} \;,
\end{equation}
and the geometry is nowhere singular for $k^2 >1$. The solution depends on two parameters, $k^2$ and $V_0$. Standard model physics is represented by $V_0$, and for every value of $V_0$ there is a solution with a flat 4D hyper-surface, i.e. the model displays self-tuning.\footnote{There is an interesting {\em brane limit} to the smooth solution (\ref{tamvakis-phi}) and (\ref{tamvakis-A}), defined as $V_0, a \rightarrow \infty$ and $k^2 \rightarrow 1$ so that $\xi^2 := 8 \kappa^2 V_0 (k^2-1)/3$ remains finite. In this limit, $A_0(y) \rightarrow -\xi |y|/4$, the scalar field and potential part of the 5D action goes to
\begin{equation}
\int_{-\infty}^{\infty} \left(\frac{1}{2} \phi'^2 + U(\phi) \right) dy \rightarrow 
\int_{-\infty}^{\infty} \left(\frac{-3\xi^2}{8\kappa^2} + \frac{ 3\xi}{2\kappa^2}\delta(y) \right) dy \;.
\end{equation}
Now it corresponds to the action of the Randall-Sundrum model \cite{randall}, with bulk cosmological constant $\Lambda=-3\kappa^{-2}\xi^2/8$ and brane tension $V_{brane}= 3\kappa^{-2}\xi/2$. Note that the RS fine-tuning relations are fulfilled automatically!}

The solutions (\ref{deltaA1p}), (\ref{deltaphi1p}) of the perturbed bulk field equations directly show the dynamical instability of the KT-model since the bulk equations are valid everywhere and we have no jump conditions for this model. Now we will discuss two special perturbation amplitudes that form part of these solutions.

First we consider a perturbation with $h_3=0$ in (\ref{deltaA1p})
\begin{equation}\label{deltaAspec1}
\delta A_L  =  A_0' h_2 t \qquad \mbox{and} \qquad \delta \phi_L=\phi_0' h_2 t \;,
\end{equation}
i.e. the model is unstable against small perturbations in the scalar field. The instability could have the unpleasant effect that the 4D Planck mass becomes time dependent, but a second order analysis is needed to settle this point. 
To first order in $\delta A$ we have
\begin{equation}\label{pert-planck} 
M_p^2 = M_5^3 \int_{-\infty}^{\infty} e^{2A(t,y)}\,  d y  \cong
M_{p0}^2 + M_5^3\int_{-\infty}^{\infty}2\,\delta A(t,y) e^{2A_0(y)}\, dy\;,
\end{equation}
where $M_{p0}$ denotes the unperturbed 4D Planck mass 
from (\ref{tamvakis-planck}). The special perturbation yields
\begin{equation}\label{pert-planck1} 
M_p^2 \cong M_{p0}^2 + M_5^3 \int_{-\infty}^{\infty}2 A_0'(y) e^{2A_0(y)} h_2 t\, dy = M_{p0}^2\;,
\end{equation}
and there is no change compared to the static solution in this approximation. It is possible that the (unknown) dynamical solution of the full field equations, which has (\ref{deltaAspec1}) as first order time expansion coefficient, also leads to usual 4D gravity on the brane. However it is not clear to us 
whether energy is still conserved on the brane in such a dynamic geometry.

Now we set $h_2=0$ in (\ref{deltaA1p}). Then the perturbation is
\begin{equation}\label{deltaAspec2}
\delta A_L  =  h_3 t \qquad \mbox{and} \qquad \delta \phi_L=0 \;.
\end{equation}
Thus there is also an instability if the scalar field $\phi$ is not perturbed. The thick brane model is unstable even under homogeneous gravitational perturbations\footnote{In the brane limit this mode survives unchanged and shows that the Randall-Sundrum-model is unstable under purely gravitational modes. This result was also found in \cite{durrer}.}. The perturbed 4D Planck mass is now time-dependent
\begin{equation}\label{pert-planck2} 
M_p^2 \cong M_{p0}^2  + M_5^3\int_{-\infty}^{\infty}2  e^{2A_0(y)} h t\, dy = M_{p0}^2 (1+2h_3 t)\;.
\end{equation}
The gravitational constant $G$ is proportional to $1/\sqrt{1+2h_3 t} \cong 1 - h_3 t$. This need not be in conflict with observation if $h_3$ is small enough, but it surely is an unpleasant effect of the instability.
Although the physical interpretation of these instabilities is not complete, 
we see that the KT self-tuning model requires a fine-tuning to be stable
in time.

\subsection{The Instability of the ADKS-KSS Model}\label{pertubation-adks}

In the ADKS-KSS self-tuning model (see \cite{adks} and \cite{kss}) the bulk potential $U(\phi)$ is zero and the static solution is 
\begin{eqnarray}
        A_0 (y) & = & \frac{1}{4} \ln \left(1- \frac{|y|}{y_c} \right) \label{A_0-adks} \\
        \phi_0 (y) &=& \phi_b + \sqrt{\frac{3}{4\kappa^2}}\ln \left(1- \frac{|y|}{y_c} \right) \label{phi_0-adks} \\
        \lambda(\phi) &=& c \; e^{- 2\kappa \phi /\sqrt{3}}  \label{lambda-adks} \\
         y_c  &:=& \frac{3 }{c} e^{- 2\kappa \phi_b /\sqrt{3}} \; .
\end{eqnarray}
In the following, we will obtain the same instability with the more general 
method of gauge invariant linear perturbation theory.

We first consider a perturbation with $h_3=0$
\begin{displaymath}
\delta A_L  =  A_0' h_{\pm} t \qquad \mbox{and} \qquad \delta \phi_L=\phi_0' h_{\pm} t \;.
\end{displaymath}
The field equations are only valid in the bulk, i.e. for $y \not= 0$. Therefore we actually have two integration constants, $h_+$ where $y>0$, and $h_-$ where $y>0$. $A_0$ and $\phi_0$ are ${\mathbb Z}_2$ symmetric, $A_0'$ and $\phi_0'$ are ${\mathbb Z}_2$ antisymmetric with a jump at $y=0$, consequently we have to set $h:=h_+=-h_-$ to get continuous perturbation amplitudes $\delta A_L$ and $\delta \phi_L$: 
\begin{eqnarray}\label{deltaAspec1-adks}
\delta A_L(t,y)  = \mbox{sgn}(y) A_0'(y) h t = \frac{-h t}{4y_c} e^{-4 A_0(y)}\\
\delta \phi_L = \mbox{sgn}(y) \phi_0'(y) h t = \frac{-h t}{y_c}\sqrt{\frac{3}{4\kappa^2}} e^{-4 A_0(y)}\ \;.
\end{eqnarray}
These perturbations also satisfy the first order jump conditions (\ref{jumpcond-A-1}) and (\ref{jumpcond-phi-1}), so they show the dynamical instability of this model against small changes in the scalar field. The solution is analogous to the dynamical solution found in \cite{binetruy} \footnote{The dynamical solutions in \cite{binetruy} are obtained with a diffeomorphism and orbifold projection from the static solution. Therefore it is possible to transform away the perturbation amplitudes locally in the bulk. The transformation, under which the metric remains diagonal, corresponds to the vector field with $X^4 = - h t\, \mbox{sgn}(y)$ and $\xi= ht \int \exp(-2A_0(y))\, \mbox{sgn}(y)\, dy$. Note that this is not a gauge transformation because $X \notin C^1$. }, the perturbations $\delta A_L$ and $\delta \phi_L$ are proportional to the first order terms in a time expansion of their dynamical solution around (\ref{A_0-adks}) and (\ref{phi_0-adks}). See \cite{binetruy} for the physical interpretation of this instability.

The homogeneous gravitational mode (\ref{deltaAspec2}) is continuous and 
fulfills the first oder jump conditions since the jumps are zero and $\delta \phi_L$ is zero as well. Therefore this model is also unstable under homogeneous gravitational perturbations. 

\section{Results and Conclusions}

Using linear perturbation theory we have demonstrated the instability of
some brane models. In particular, we have reproduced the instability of the ADKS-KSS model \cite{adks,kss} obtained by a diffeomorphism and orbifolding in 
\cite{binetruy}. Linear perturbation theory also works for the smooth KT brane model \cite{tamvakis} and presumably also for brane world models with non-trivial bulk potential.

The KT self-tuning model is not stable against small perturbations of the scalar field and it is even unstable under a purely gravitational mode since there is a metric perturbation linear in time which solves the field equations with the unperturbed scalar field. This mode causes the 4D effective Planck mass to be time dependent. The physical interpretation of these instabilities is not clear to us; it is possible that they affect energy conservation on the brane, 
as in the ADKS-KSS model. The instabilities show that the KT self-tuning model requires a fine-tuning to obtain the static solution.

\section{Acknowledgements}

We wish to thank Norbert Straumann, Ruth Durrer, Timon Boehm and Tobias Kaufmann for useful discussions and comments.

\pagebreak

\end{document}